# NUMERICAL STUDYING MECHANICS OF A STATIONARY RANGE FORMATION IN THE WIND-WAVE SPECTRUM

**Vladislav G. Polnikov[1], Fangli Qiao[2,3] and Jing Lu[2]**

[1]A.M. Obukhov Institute of Atmospheric Physics of RAS, Moscow, Russia

[2]First Institute of Oceanography of SOA, Qingdao, China

[3]Laboratory for Regional Oceanography and Numerical Modeling, Qingdao National Laboratory for Marine Science and Technology, Qingdao, China

The process of a stationary range formation in the wind-wave spectrum is investigated numerically. The evolution equation for the two-dimensional wind-wave spectrum is numerically solved by using an exact calculation of the Hasselmann's kinetic integral with exploring several parametrizations for the wave-pumping and wave-dissipation mechanisms. The following results are established. First, there is no any inertial interval in the spectral frequency band of real wind waves. Therefore, there is no reason for the Kolmogorov-type spectra formation in this case. Second, the ratio between the input and dissipation mechanisms is responsible for a stationary range formation in the wind-wave spectrum. Third, this ratio enables us to establish all known forms for a stationary range of the spectrum, if proper mathematical representations for the input and dissipation mechanisms are chosen.

## 1. Introduction. Surface wave spectral forms

The point of understanding mechanism of a stationary range (or equilibrium range) formation in the frequency spectrum for wind waves (shortly - "the tail" of spectrum) is one of traditional problem in the wind-wave physics. Its importance is stipulated by the fact that a shape of the spectrum tail is provided by a balance between evolution mechanisms for wind waves. Therefore, robust measurements of the spectrum-tail shape and theoretical interpretation of its formation provide us a criteria for understanding evolution mechanisms of wind-waves and formation of their spectrum is correct or not.

The beginning of studying this issue was initiated by Phillips (1958) nearly 60 years ago, which was based on a simple consideration of dimensions. Phillips has proposed the so-called "equilibrium range" of the frequency spectrum for wind waves, $S(\omega)$, in the form

$$S(\omega) = \alpha_F g^2 \omega^{-5} \qquad . \tag{1}$$

Here, $S(\omega)$ is the energy spectrum for a water-surface elevation, given in units $L^2T$; $\omega$ is the circular frequency (rad/s), $g$ is the acceleration due to gravity, and $\alpha_F$ is the so-called "Phillips' constant". Form (1) was proposed to be valid at the tail of spectrum, i.e. far from the peak frequency: at least for $\omega > 2\omega_p$ (Hasselmann et al. 1973; Liu 1989), where $\omega_p$ is the peak frequency corresponding to the peak of spectrum $S(\omega)$. Derivation of formula (1) needs no assumption about mathematical representations for the wind-wave evolution mechanisms.

The justification of form (1) is based only on the formal assumption that the shape of equilibrium and intensity-saturated part of the wind-wave spectrum does not depend on the wind speed, $U$, causing the waves. Due to this, sometimes spectrum (1) is called as "the saturation spectrum", to separate it from "the equilibrium spectrum" depending on a wind value. In fact, the equilibrium and saturation ranges are the parts of a stationary tail in the wind-wave spectrum (Phillips 1985; Donelan et al. 1985; Resio et al. 2004; Tamura et al. 2014; among others). For this reason, in this paper we prefer to use the generalized term: "the stationary range" of spectrum, do not separating terms of the equilibrium and saturation spectrum, often used in literature (e.g., Tamura et al, 2014).

Spectral shape (1) has been repeatedly confirmed in numerous field observations (for example, Pierson & Moskowitz 1964; Hasselmann et al. 1973; Babanin & Soloviev 1998; Tamura et al. 2014). Usually, it is observed for a spatially homogeneous and constant wind of various values. Herewith, it was established that a magnitude of $\alpha_F$ depends on the wave-formation conditions (for example, Hasselmann et al. 1973; Babanin & Soloviev 1998). But this fact is not principal for our consideration.

In the 60s of the 20$^{th}$ century, other theoretical models dealing with the wind-wave spectrum tail were appeared. First of all, Kitaigorodskii (1962) has proposed an equilibrium range in the form

$$S(\omega) \propto U_{10} g \omega^{-4} \qquad . \tag{2}$$

Here, the intensity of spectrum tail depends on the wind speed at a standard horizon, e.g., $U_{10}$ (or on the friction velocity, $u_*$, related to $U_{10}$ by ratio, $u_* = C_d^{1/2} U_{10}$, where $C_d$ is the empirical drag coefficient). Kitaigorodskii (1962) has found form (2) by means of dimensional consideration. Herewith, he has assumed that the rear part of wind-wave spectrum is formed due to the Kolmogorov's flux of energy, $\varepsilon$, toward to the higher frequencies. However, it was only a hypothesis. Later, making several suggestions aimed to justify his hypothesis, Kitaigorodskii (1983) has estimated the specially defined energy flux, $\varepsilon$, and found that $\varepsilon \propto (U_{10})^3$. Thus, form (2) was transformed to the representation

$$S(\omega) \propto \varepsilon^{1/3} \omega^{-4} \tag{3}$$

corresponding to form (2) as the certain Kolmogorov-type spectrum.

To separate form (2) from the Phillip's form (1), the former was called as "the equilibrium spectrum", taking in mind that the equilibrium is given by some kind balance between the wind-wave evolution mechanisms, for which the nonlinear one plays the crucial role in the spectrum-shape formation.

In parallel, Zakharov & Filonenko (1966) derived analytically nearly the same form,

$$S_Z(\omega) \propto \omega^{-4}, \tag{4}$$

as the exact stationary solution of the four-wave kinetic equation of the kind

$$\partial S(\omega, \theta)/\partial t = I_{NL}[S(\omega, \theta)], \tag{5}$$

where $I_{NL}[S]$ is the conservative four-wave kinetic integral, in the **k**-space having the kind

$$I_{NL}[S(\mathbf{k})] = = 4\pi \iiint M^2_{\mathbf{k},\mathbf{k}_1,\mathbf{k}_2,\mathbf{k}_3} F3(S_{\mathbf{k}}, S_{\mathbf{k}_1}, S_{\mathbf{k}_2}, S_{\mathbf{k}_3}) \delta(\mathbf{k}+\mathbf{k}_1-\mathbf{k}_2-\mathbf{k}_3)\delta(\omega+\omega_1-\omega_2-\omega_3) d\mathbf{k}_1 d\mathbf{k}_2 d\mathbf{k}_3$$

derived for the first time in (Hasselmann 1962). Here, $M_{\mathbf{k},\mathbf{k1},\mathbf{k2},\mathbf{k3}}$ is the four-wave interacting matrix, $F3(S_{\mathbf{k}}, S_{\mathbf{k1}}, S_{\mathbf{k2}}, S_{\mathbf{k3}})$ is the special cubic form of the four wave spectra, $S_{\mathbf{k}i}$ ($i$=0,1,2,3, the zero subindex is omitted in (5a)), and $\delta$ is the Dirac's delta-function. As the form of kinetic integral $I_{NL}[S]$ is well-known (e.g., Hasselmann 1962; Zakharov & Zaslavskii, 1982; Komen et al. 1994; The WISE Group 2007; van Vledder 2006), here we restrict ourselves by the symbolic writing of the integral. The explicit form of $I_{NL}[S]$ will not be needed farther.

Stating that two-dimensional spectrum $S(\omega, \theta)$ has the power-like form, $S(\omega, \theta) \propto \omega^{-n}$, on the whole frequency band, from 0 to $\infty$, and spread isotropically over the angle, $\theta$, Zakharov & Filonenko (1966) have found that spectrum $S_Z(\omega)$ of form (4) does put kinetic integral to zero: $I_{NL}[S_Z(\omega, \theta)]=0$. The authors identified form (4) as the Kolmogorov's spectrum of the constant energy-flux, $\varepsilon$, to the higher frequencies, provided by the energy-source, *In*, and energy-sink, *Dis*, located at the zero and infinity points on the frequency band, respectively. Thereby, in addition to the Kitaigorodskii's approach, Zakharov & Filonenko have specified the mechanism determining the flux $\varepsilon$, namely: the four-wave nonlinear interactions in waves. This specification was explicitly formulated by Zakharov & Zaslavskii (1982) just in form (3), conforming the Kitagorodskii's hypothesis. Moreover, developing an analytical approach to solving kinetic equation (5), Zakharov & Zaslavskii (1982) have found another kind of the stationary Kolmogorov-type solution of the form

$$S(\omega) \propto \phi^{1/3} \omega^{-11/3}, \qquad (6)$$

corresponding to the constant wave-action flux, $\phi$, towards to the lower frequencies. By the way, the same result was established in (Kitaigorodskii 1983) by means of dimensional consideration.

For the first time, the experimental confirmation of the wind-wave spectrum of form (2) was appeared in the tank experiments of Toba (1972), who has represented the found spectrum in the form

$$S(\omega) = \alpha_T u_* g \omega^{-4}, \qquad (7)$$

where $\alpha_T$ is the dimensionless constant. Despite of similarity between forms (2) and (7), the spectrum of form (7) is called as "the Toba's spectrum", and $\alpha_T$ does "the Toba's constant". The features of this spectrum were studied by a lot of authors cited, for example, in (Babanin & Soloviev 1998; Resio et al. 2004; Tamura et al. 2014). They are out of our consideration.

After that, the Phillips' spectrum was revised in favor of the Kitaigorodskii-Zakharov-Toba's results in a huge series of theoretical works including (Kitaigorodskii et al. 1975; Kitaigorodskii 1983; Zakharov & Zaslavskii 1982; Phillips 1985; among others).

Since late 80s of the 20$^{th}$ century, all the mentioned forms of the stationary part of the wind-wave spectra were observed experimentally (see references in Komen et al. 1994; The WISE Group 2007; Resio et al. 2004; Tamura et al. 2014; and cited therein). Regarding to the analysis of observations, the field measurements data of JONSWAP (Hasselmann et al. 1973) were also revised (Kahma & Calkoen 1992) in favor of forms (2) and (7). In addition to this, there were executed several rigorous analysis of wave measurements in open water reservoirs (see, Donelan et al. 1985; Liu 1989; Resio et al. 2004; Tamura et al. 2014, among numerous others), confirming the existence of forms (1), (2) and (7) in field waves. On the other hand, some special studies of the stationary part of spectrum for the natural wind waves (Liu 1989; Rodrigues & Soares 1999; Young 1998) have shown that a decay law of the spectrum tail can vary within very wide limits: from $\omega^{-3}$ to $\omega^{-7}$.

All of these observed spectral shapes need their own treatment.

**2. Treatments of the spectral forms and their analysis**

At present, spectral forms (2, 4, 7), obtained in experiments, are interpreted as the Kolmogorov-type spectra caused by the nonlinear interactions in waves (Phillips 1985; Donelan et al. 1985; Kahma & Calkoen 1992; Resio et al. 2004; Tamura et al. 2014; among others). The justification for these treatments is based on the mentioned theoretical results by Kitaigorodskii, and Zakharov with coauthors. In this regard, the recent experimental work by Tamura et al. (2014) is the most instructive, in which the both spectral shapes of forms (2) and (1) were obtained. Working in the wave-vectors **k**-space, they have found that "the saturation spectrum, $B(k) = k^3 S(k)$, is linearly increased with $U_{10}$ at the range 1<k<5 rad/m, when wind speed was less than 10 m/s; and $B(k)$ is tended to be saturated under higher winds. In the intermediate range (5 < k < 10), $B(k)$ showed a weak dependence on $U_{10}$ over the entire wind speed range" (citation from Tamura et al, 2014).

Developing methods for interpreting observations, Tamura et al, (2014) have performed a number of simulations describing a stationary-part formation in the wind-wave spectrum, $S(\mathbf{k})$, given in the wave-vector **k**-space. To this aim, they have used the new version of numerical model WAVEWATCH (WW), initially proposed by Tolman & Chalikov (1996), what allows them to test the influence of input and dissipation functions, $In(\mathbf{k},S)$ and $Dis(\mathbf{k},S)$, together with the exact calculation of kinetic integral $I_{NL}(\mathbf{k},S)$, installed into the model. All these terms were represented in the **k**-space.

As a result of modeling, they have obtained the tail of spectrum consisting of two parts: one of form (1) and the other of form (2). Carrying out some, rather original estimates of the evolution terms $I_{NL}(\mathbf{k},S)$, $In(\mathbf{k},S)$ and $Dis(\mathbf{k},S)$, integrated though various ranges of wave-vector **k**, Tamura et al. (2014) have made the conclusion that the nonlinear mechanism is prevailing in the spectral range of form (2). Therefore, referring to Resio et al. (2004), the authors concluded that the observed spectral form (2) is the Kolmogorov's spectrum of the energy-flux upward in the wave numbers. However, Tamura et al. (2014) did not explicitly show an existence of any inertial interval in the range of spectral form (2), and did not show any separation between source and sink terms in the wave-number domain.

Regarding to spectral form (1), it is traditionally interpreted by all the mentioned authors as the Phillips' spectrum provided by the hydrodynamic instability and total dissipation of the wave energy exceeding its saturation level (see remark on the terminology in the introduction). We note, however, that in numerical simulations of wind waves, such a part of the spectrum is usually prescribed in advance (Tolman at al. 2014), since it is difficult to achieve numerically a complete balance of evolution terms $I_{NL}(\mathbf{k},S)$, $In(\mathbf{k},S)$ and $Dis(\mathbf{k},S)$ at the tail of spectrum, in practical models of wind waves. Therefore, to provide the necessary law of spectrum decay by numerical wind-wave models, some implicit measures of smoothing are used, prescribing the tail of spectrum or introducing the so-called "limiters" (The WAMDI group 1988; Tolman at al 2014). Choosing one or another limiter, one can get the desired prescribed form of spectrum.

As to the spectrum of form (2), for a convincing proof of the Kolmogorov's spectrum existence under natural conditions, an explicit demonstration of the inertial-interval existence for natural wind-waves is necessary, at least. No such evidence was presented in all the works mentioned above. Though, it is well known that a presence of the inertial interval is one of the postulates of the Kolmogorov's theory, which ensures a possibility to have a constant energy (or action) flux through the spectrum (see, for example, Monin & Yaglom 1971). Otherwise, it is necessary to seek for another explanation on the observed spectral shape of form (2).

Some alternative interpretation of spectral shapes (2) and (4) is available at present. It is based on the fact that spectra of form (4) can be a simple consequence of the impact on the liquid surface by the random fluctuations of air pressure, $P(t)$, having the white-noise spectrum, $S_P(\omega) = const$ (Polnikov & Uma 2014). In the latter paper, basing on the Euler's equations for a wavy fluid surface, it was shown that the wave spectrum could have form (4) even in the linear approximation (if the nonlinear and dissipation terms are neglected).

At the same time, there are clear numerical proofs of theoretical possibility for existing spectral forms (4) and (6) as the Kolmogorov's spectra. Indeed, the direct numerical solutions of the generalized kinetic equation of the form

$$\partial S(\omega,\theta)/\partial t = I_{NL}[S(\omega,\theta)] + In(\omega,\theta) - Dis(\omega,\theta), \qquad (8)$$

which was performed for the first time by the leading author (Polnikov 1994, for isotropic spectrum; Polnikov 2001, for anisotropic one), have shown that the Kolmogorov's spectra of forms (4) and (6) are actually realized due to the four-wave nonlinear interactions in waves. Later, the similar results were obtained by other authors (for example, see, Badulin et al. 2005). However, this fact requires a presence of energy source $In(\omega,\theta)$ and sink $Dis(\omega,\theta)$, spaced apart each from other in the frequency band. It means an explicit presence of the inertial interval in the frequency band, in which the nonlinear term, $I_{NL}[S]$, "works" alone, as required by the theory (Monin & Yaglom, 1971). Though, under the natural conditions, when energy input $In(\omega)$ and wave dissipation $Dis(\omega)$ are distributed over the entire frequency band, $0 < \omega < \infty$, where a wave spectrum $S(\omega)$ exists, there is not any explicit proof of the inertial interval existence, as it was already noted.

For the representations of $In(\omega)$ and $Dis(\omega)$, known from the literature, the verification of possibility for the inertial-interval existence can be easily carried out numerically. For example, in the case of source and sink functions $In(\omega)$ and $Dis(\omega)$ taken from the numerical wave model WAM (The WAMDI group 1988), Polnikov & Uma (2014) have shown, for one particular time moment of wind-wave evolution, that the inertial interval is absent in real wind waves. In the same paper, it was also shown that the traditional representation spectra in forms $S(\omega) \propto \omega^{-4}$ and $S(\omega) \propto \omega^{-11/3}$, given on a limited frequency band, do not turn kinetic integral $I_{NL}[S]$ to zero. These facts once again make it arguable a using the Kolmogorov's model for interpretation spectral forms (4) and (6) in real wind waves. Therefore, in the case of real wind waves, this point requires its further clarification.

In addition to that, there are several special studies in which they have shown that a decay law of the spectrum-tail can vary within very wide limits: from $\omega^{-3}$ to $\omega^{-7}$, in the stationary range of natural wind-waves spectra (for example, Liu 1989; Rodrigues & Soares 1999, Young 1998). These results also have not a clear theoretical explanation in terms of nonlinear interactions, proposed in papers by Zakharov and coauthors mentioned above.

In this regard, it is worthwhile to dwell on the recent paper by Zakharov & Badulin (2011), where they have tried to prove that the spectrum tail should be predominantly formed by the nonlinear mechanism. Their theoretical justification is based on the statement that the nonlinear term, $I_{NL}[S]$, standing in the r.h.s. of Eq. (8), should suppress the impact of the other terms, $In(S)$ and $Dis(S)$, in the tail range of wind-wave spectrum. To justify this, they proposed to be share nonlinear term $I_{NL}[S]$ on two parts. In the one-dimensional representation, $I_{NL1}(\omega)$, it is written as:

$$I_{NL1}(\omega) \equiv Nl(\omega) = Nl^{+}(\omega) + Nl^{-}(\omega) \qquad . \qquad (9)$$

Here, $Nl^{+}$ is the positive part, and $Nl^{-}$ is the negative part of the total nonlinear term, $Nl(\omega)$. Then, attracting their own exact calculations of terms $Nl^{+}(\omega)$ and $Nl^{-}(\omega)$ and several known parametrizations for $In(\omega)$ and $Dis(\omega)$, Zakharov & Badulin (2011) have shown that in the frequency range $\omega > \omega_p$, for all used wind-wave spectra shapes, the following ratios take place:

$$Nl^{+}(\omega) >> In(\omega) \qquad \text{and} \qquad |Nl^{-}(\omega)| >> |Dis(\omega)| \qquad . \qquad (10)$$

In their opinion, ratios (10) prove the fact of existence of inertial interval in real wind waves, providing the basis for the Kolmogorov-type spectra formation in this case.

However, the approach proposed by Zakharov & Badulin (2011), in our opinion, is incorrect from both the mathematical and the physical point of view. First, the algebraically similar terms, $Nl^{+}(\omega)$ and $Nl^{-}(\omega)$, each of which is cubic in spectrum $S$, in contrary, must be summed before their comparison with other terms in the r.h.s. of Eq. (8). Second, it is physically incorrect to share a single mechanism on two ones. There are no independent mechanisms described by $Nl^{+}(\omega)$ and $Nl^{-}(\omega)$. It is the mathematical representation of nonlinear term $Nl(\omega)$, only.

Although integral $I_{NL}[S]$ can be represented mathematically as a sum of two terms with the opposite signs, in reality it is a single mechanism. Thus, only the total value of nonlinear term, $Nl(\omega)$, must be compared with the total sum of the residual terms in the r.h.s. of equation (8),

$$B(S) = In(S) - Dis(S). \tag{11}$$

Later, *B(S)* is called as the balance function. This quantity, in fact, describes the combined "source-sink-mechanism" balancing the wave energy. This mechanism relates the wind-waves system to the "external" air-water system. It exists on the background of the "internal" evolution mechanism, the conservative four-wave nonlinear interactions. (By the way, these source and sink mechanisms are physically independent, and exist in parallel).

Note that in the most cases, the point of applicability of the Kolmogorov's model to a wind-wave spectrum was being studied, mainly, by analytical methods (Kitaigorodskii 1983; Zakharov & Filonenko 1966; Phillips 1985; Zakharov & Zaslavsky 1982), attracting certain simplifications, approximations, and hypotheses, but avoiding the point of proving the inertial-interval presence. In the rare cases of using numerical methods (Zakharov & Badulin 2011; Tamura at el. 2014), these authors also did avoid the issue of proving an existence of the inertial interval in the case of real wind-wave spectra, did not trying to reproduce the form of the "external" balance function, *B(ω) = In(ω)-Dis(ω)*. It seems that such an approach is due to a less justification of parameterizations for *In(ω)* and *Dis(ω)* in comparison with the exact term, *Nl(ω)*.

However, at present, the numerical simulation of wind waves has already reached a level when its accuracy approaches to the one of direct buoy-measurements of waves (The WISE Group 2007; Samiksha et al. 2015), and even exceeds the direct satellite measurements of wave heights (Chen-Zhang et al. 2016; Kubryakov et al.,2016). This means that understanding the physics of all the mechanisms for wind-wave evolution has reached a significant progress, including several justified analytical parameterizations for functions *In(S)* and *Dis(S)*. In addition to this, there are several exact algorithms for calculating kinetic integral $I_{NL}[S]$ and solving equation (8) (for example, Masuda 1980, Polnikov 1989, 1990; Resio & Perrie 1991; Komatsu & Masuda 1996; van Vledder 2006). This allows executing a direct numerical simulation of the wind-waves spectrum evolution and clarifying both the point of the inertial-interval existence and the question about mechanisms which are responsible for a formation the tail of wind-wave spectrum. By this way, one can also clarify the point of correctness of sharing integral $I_{NL}[S]$ on two terms. To do this, it is sufficient to perform numerical solution for Eq. (8), using the exact calculation of $I_{NL}[S]$ and exploring several mathematical representations for the source and sink functions, *In(S)* and *Dis(S)*, known from the practice of numerical simulation of waves.

The present work is devoted to solution of the mentioned physical issues by means of numerical experiments.

3. **The essence of the problem**

The problem to be solved is the following. As it is well known, the evolution of the two-dimensional energy spectrum of wind waves, $S(\omega, \theta; t, x, y)$, is described by the generalized kinetic equation having, in the simplest case, the form (Komen et al. 1994)

$$\frac{dS}{dt} \equiv \frac{\partial S}{\partial t} + C_{gx}\frac{\partial S}{\partial x} + C_{gy}\frac{\partial S}{\partial y} = F \equiv In(S,U) + Nl(S) - Dis(S,U). \qquad (12)$$

In Eq. (12), the l.h.s. means the full derivative of the spectrum with respect to time, and the r.h.s. is the so-called source(better, forcing) function $F$ (hereinafter - SF). The l.h.s. of (12) is responsible for the mathematical part of numerical model for wind waves. It should be complemented by the boundary and initial conditions, and the input wind-field, $\mathbf{U}(t, x, y)$. The SF contains a physical content of the model. It includes three basic mechanisms of evolution: (a) the mechanism of wave-energy exchange with the wind ("input" or "pumping" function), $In(S,U)$; (b) the conservative mechanism of nonlinear interactions between waves ("nonlinearity"), $Nl(S)$; (c) the mechanism of wave-energy losses ("dissipation"), $Dis(S,U)$.

Each of these SF-terms, principally, should be derived from the basic equations of the wind-wave dynamics (Komen et al. 1994; The WISE group 2007; Polnikov 2010). However, even in the simplest case of ideal fluid, these dynamics equations (e.g., the Euler's equations) are not amenable to an analytical transform to Eq. (12). Nevertheless, by using a number of simplifying assumptions, methods of Fourier expansions, and perturbation theory on a small parameter, one can obtain expected analytical forms for SF-terms. Though, the final exact forms for SF-terms cannot be analytically obtained, and they are to be semi-empirically specified. The only exception is the nonlinear term, $Nl(S)$ (Hasselmann 1962), repeatedly mentioned above. Finally, the intermediate theoretical results for terms $In$ and $Dis$ provided, for example, by the Miles' model for the input-term (Miles 1957), or by the models for the dissipation-term (Hasselmann 1974; Polnikov 2012), require a search for adequate analytical approximations, and subsequent tuning and verification of them (Komen et al. 1994; The WISE group 2007).

The dominance of nonlinear mechanism in a wind-wave evolution has been pointed by numerous researchers (Hasselmann et al. 1973; Komen et al. 1984; Young & van Vledder 1993; Badulin et al. 2005; Zakharov & Badulin 2011; among others). Though, the actual role of $Nl$-term cannot be assessed theoretically without consideration the roles of other evolution mechanisms. This point should be solved by numerical simulations, realized with physically correct and numerically justified parameterizations of source terms and solution of equation (12).

Our experience in exact calculations of $Nl$-term(Polnikov 1989; Polnikov & Uma 2014), and in the wave modeling and SF-terms verification (for example, Polnikov & Innocentini 2008; Samiksha et al. 2015), does urge us to doubt that the nonlinear mechanism is the only one, which is responsible for the equilibrium-tail formation in the spectrum of real wind waves. It seems that a final decision of this point may be found by numerical simulations, only.

As noted above, at present there are a huge number of verified parametrizations for *In* and *Dis* (Komen et al. 1994; The WISE Group 2007, among others), which allow us to solve the mentioned problem numerically by means of varying expressions for *In* and *Dis*. Herewith, the nonlinear mechanism, *Nl(S)*, should be setup by its exact numerical representation, to be in accordance with the statements of paper (Zakharov & Badulin, 2011). By this way, first of all, one could check, if the inertial interval takes place on the frequency band in the case of real input and dissipation terms. Second, by numerical calculations it is easy to clarify an impact of all the evolution mechanisms on the formation a decay law of wind-wave spectrum. The proper criteria is very simple: if any variations of the frequency dependences for *In(ω)* and *Dis(ω)* have no influence on the shape of spectrum tail, then the spectrum tail is fully formed by the *Nl*-mechanism, and vice versa. If an increase or decrease of the frequency dependence of functions *In (ω)* or *Dis (ω)* leads to a noticeable changing a value of falling-law parameter *n* in the spectral form

$$S(\omega) \propto \omega^{-n}, \tag{13}$$

then the *Nl*-mechanism is not the only one, which is essential in formation the shape of spectrum tail.

In such numerical experiments, the physical justification of representations for *In(ω)* and *Dis(ω)* is not so important. Therefore, any kind of known forms for them could be used. The only important point is that the nonlinear term must be taken in the exact representation.

The essence of this work is to clarify all the mentioned points numerically.

4. **The method of research and formulation of the numerical task**

4.1. *The evolution mechanisms representations*

4.1.1 Nonlinear term.

To solve the task posed, we should, first, to demonstrate effectiveness of the *Nl*-term calculation and kinetic equation solution, and second, to specify analytical representations for the other SF-terms to be used. We begin with term *Nl(S)*, as the most certain one, which is well-studied analytically (Zakharov & Filonenko 1996; Zakharov & Zaslavskii, 1982) and numerically (Masuda 1980; Polnikov 1989; Badulin et al, 2005; van Vledder 2006, among others). Here we use the algorithms derived in Polnikov (1989, 1990).

Effectiveness of our *Nl*-term calculation and kinetic equation solution is shown in the plots for a certain case of numerical solution of the pure nonlinear kinetic equation (5) (figures 1a,b).

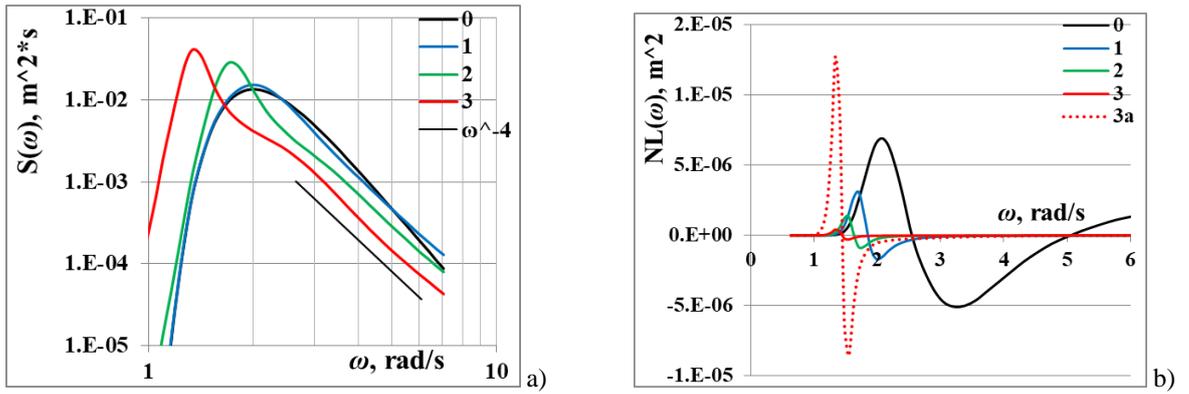

FIGURE 1. a) Evolution of spectrum $S(\omega,t)$, as an exact numerical solution of kinetic Eq. (5). Initial shape of $S(\omega,\theta)$ is the Pierson-Moskovitz spectrum with $\omega_p=2$rad/s and the angular distribution $\cos^2(\theta)$. b) Change of nonlinear transfer $Nl(\omega,t)$ corresponding to an exact numerical solution of kinetic Eq. (5). Line 0 corresponds to time $t = 0$, line 1 to $t = 1.8*10^3$ s, line 2 to $t = 3.1*10^4$ s, line 3 to $t = 3.8*10^5$ s. The line $\omega^{-4}$ is shown below in a). Line 3a in b) is line 3 multiplied by 30.

According to the well-recognized features of such solution, the shape of spectrum $S(\omega,\theta)$ (and the shape of $S(\omega)$, consequently) becomes self-consistent, not depending of the initial form of spectrum $S(\omega,\theta)$ (Polnikov 1990; Komatsu & Masuda 1996; Badulin et al, 2005). For future consideration, it is essential to note the following two important features of the solution: (a) the falling law for the tail of 1D-spectrum $S(\omega)$ becomes very close to law "$\omega^{-4}$", and (b) the 1D nonlinear transfer $Nl(\omega)$ becomes very small (retaining the negative value) in the range $\omega > 2\omega_p$. To show this more clearly, special line 3a is plotted in figure 1b.

These features are in good accordance with the theory (Zakharov & Filonenko 1966), despite of the fact that spectrum $S(\omega,\theta,t)$ is not stationary, anisotropic in angular, and permanently shifting to the lower frequencies. We do not dwell here on these details more, left it to a separate consideration, including an analysis of the 2D-spectrum shape. But we keep features (a) and (b) in mind, to compare them with their analogues in the case of the generalized kinetic-equation solution.

Regarding to the choice of representations for terms *In* and *Dis*, there is a considerable arbitrariness of it, due to the lack of exact theories, as mentioned above. We start from the simplest representations for terms *In* and *Dis*, used in the widely-spread European model WAM (The WANDI group 1988; Komen et al 1994). Then, we use a little more complicated forms for *In* and *Dis* terms, used in the Chinese model MASNUM (Yuan et al. 1991: Yang at al. 2005), widely tested and effectively used even in surface wave-circulation coupling practice (Qiao et al. 2004). And finally we use more complicated representations for *In* and *Dis* terms, proposed in (Polnikov 2005). They have been successfully verified in numerical simulations in the shells of models WAM and WW for numerous real wind-wave fields (for example, Polnikov & Innocentini 2008; Samiksha et al. 2015).

It seems that these examples for *In* and *Dis* terms are sufficient to make certain conclusions about the questions posed above.

### 4.1.2. WAM's *In* and *Dis* terms

Following to the technical note (Tolman 2014), we restrict ourselves with the simplest version: WAM-cycle3. In this case, the basic forms for *In* and *Dis*, named as $In_W$ and $Dis_W$, written in the $(\omega,\theta)$-variables and in terms of the energy spectrum, $S(\omega,\theta)$, are as follows

$$In_W(u_*,\omega,\theta,S) = C_{in}\frac{\rho_a}{\rho_w}max\left[0,\left(\frac{28u_*\omega}{g}cos(\theta-\theta_u)-1\right)\right]\omega S(\omega,\theta), \tag{14}$$

where $\rho_a$ and $\rho_w$ is the density of the air and water, respectively; $u_*$ is the friction velocity; and

$$Dis_W(u_*\omega,\theta,S) = C_{dis}(\omega/\overline{\omega})^2(\overline{\alpha}/\alpha_{PM})^2\overline{\omega}S(\omega,\theta), \tag{15}$$

where the bar over variables means the specially defined averaging over the spectrum, and $\alpha$ is the second power of wave steepness (for details, see The WAMDI group 1988; Tolman 2014). The default fitting constants are as follows: $C_{in} = 0.25$, $C_{dis} = 2.36 \cdot 10^{-5}$, and $\alpha_{PM} = 3.02 \cdot 10^{-3}$.

Everywhere, transition from the wind velocity at the standard horizon, $U_{10}$, to the friction velocity, $u_*$, is carried out by the simplest formula

$$u_* = U_{10}/28, \tag{16}$$

widely used for testing simulations (Komen et al. 1994). With no limits of generality, the value $U_{10}$ is taken to be equal to 10 m/s, and the wind direction, $\theta_u$, is assumed to be equal to 0.

The evolution equation (5) is solved with these $In_W$ and $Dis_W$ representations and with some of their modifications related to increasing frequency-dependence of function $In_W$ or $Dis_W$ (see below).

### 4.1.3. MASNUM's *In* and *Dis* terms

The MASNUM's source function is written in the **k**-space representation. For the simplest case of deep water without current, it has the typical form as the r.h.s. of (12). In the $(\omega,\theta)$-space, the input term, $In_M$, has the kind

$$In_M(u_*,\omega,\theta,S) = \alpha_M + \beta_M \omega S(\omega,\theta) \tag{17a}$$

where

$$\alpha_M = 80\left(\frac{\rho_a}{\rho_w}\right)^2 \frac{u_*^4}{g^2} \cos^4(\theta - \theta_u) H[\cos(\theta - \theta_u)] \qquad . \tag{17b}$$

Here, H[.] is the Heaviside function, $\frac{\rho_a}{\rho_w} = 1.25 \cdot 10^{-3}$, and $\beta_M$ is the same as in the WAM model (14). The *Dis*-term, $Dis_M$, is also very similar to formula (15), but with some modifications

$$Dis_M(u_*,\omega,\theta,S) = d_1 (\omega/\overline{\omega})^2 (\overline{\alpha}/\alpha_{PM})^{1/2} * \exp[-d_2(1-\varepsilon)^2 * (\alpha_{PM}/\overline{\alpha})]\overline{\omega}S(\omega,\theta) \tag{18}$$

with fitting constants: $d_1 = 1.32 \cdot 10^{-4}$, $d_2 = 2.61$, and $\varepsilon = 0.6$. Regarding to MASNUM, the evolution equation (5) is solved with these $In_M$ and $Dis_M$, only.

### 4.1.4. Polnikov's *In* and *Dis* terms

In this case, we use the well-known, linear in spectrum *S* representation for *In*-term (Miles 1957), which can be written as

$$In = C_{in}\beta(\omega, \theta, \mathbf{U})\omega S(\omega,\theta) \qquad . \tag{19a}$$

The theoretically undefined increment of growth rate, $\beta(\omega,\theta,\mathbf{U})$, is given by the semi-empirical approach (Polnikov 2005), based on paper (Yan, 1987):

$$\beta(\omega,\theta,\mathbf{U}) = \max\left\{-b_L, \left[0.04\left(\frac{u_*\omega}{g}\right)^2 + 0.00544\frac{u_*\omega}{g} + 0.000055\right]\cos(\theta - \theta_u) - 0.00031\right\}. \tag{19b}$$

With fitting constants $C_{in} = 2\pi*0.5$ and $b_L = 5 \cdot 10^{-6}$ (according to Samiksha et al. 2015), the expression for function $In(\omega,S,\mathbf{U})$ given by formulas (19a, b) is considered as the "basic" one, $In_b$.

Farther, we shall need of variations for function $In(\omega,S,\mathbf{U})$, associated with a change of its frequency dependence. In our numerical experiments, these variations will differ from basic expression $In_b$ by the factor of dimensionless frequency, $u_*\omega/g$ (see below).

As to the initial dissipation function, $Dis(\omega,S,\mathbf{U})$, we prefer to use the quadratic in spectrum *S* representation, which was theoretically justified in (Polnikov 2012) from the first principles. In the detailed semi-phenomenological form, it reads

$$Dis(\omega,\theta,S,\mathbf{U}) = C_{dis}c(\omega,\theta,\omega_p)T(\omega,\theta,\theta_u)M(\omega,\theta,\mathbf{U})\frac{\omega^6}{g^2}S^2(\omega,\theta). \tag{20a}$$

Here the function

$$M(\omega,\theta,\mathbf{U}) = \max[\beta_{dis}, \beta(\omega,\theta,\mathbf{U})] \tag{20b}$$

contains the direct wind dependence for $Dis(\omega,S,\mathbf{U})$, where function $\beta(\omega,\theta,\mathbf{U})$ is given by Eq. (19b). The fitting factor, $c(\omega,\theta,\omega_p)$, describing the dissipation intensity near peak frequency $\omega_p$, is given by the formula,

$$c(\omega,\theta,\omega_p) = 32\,max[0,(1-0.3\omega_p/\omega)] \quad . \tag{20c}$$

The angular dependent factor, $T(\omega,\theta,\theta_u)$ is given by the rations

$$T(\omega,\theta,\theta_u) = T_0 + T(\omega)T(\theta,\theta_u) \quad , \tag{20d}$$

where we use the following specification:

$$T_0 = 1 \, , \; T(\omega) = 4(\omega/\omega_p) \, , \text{ and } \; T(\theta,\theta_u) = sin^2((\theta-\theta_u)/2)\,max[1,1-cos(\theta-\theta_u)]. \tag{20e}$$

All these parameters have their own physical meanings (for details, see Polnikov 2012), though details of representations (20a, b, c, d, e) are not significant here. With fitting constants $C_{dis} = (2\pi)^2*0.4$ and $\beta_{dis} = 5\cdot10^{-5}$ (according to Samiksha et al. 2015), the expression for $Dis(\omega,S,\mathbf{U})$ given by formulas (20a, b, c, d, e) is considered as the "basic" one, $Dis_b$.

Farther variations of function $Dis(\omega,S,\mathbf{U})$, associated with a change of its frequency dependence $Dis(\omega)$, will differ from basic expression $Dis_b$ by changing factors $T(\sigma,\theta,\theta_u)$ and $M(\omega,\theta,\mathbf{U})$ (see below). All variations of $Dis(\omega,S,\mathbf{U})$ are related to an increasing or decreasing a power for its frequency dependence.

### 4.2. *Algorithm of executing the simulation task*

The main goal of our calculations is to investigate numerically the role of each evolution mechanisms, *Nl, In* and *Dis*, in formation of the wind-wave spectrum tail, by estimating their effect on the value of the decay-law parameter, *n*, given in (13). Variations of functions *In* and *Dis* concern only their mathematical representation, without any relation to their physical content. This approach is widely used in numerical experiments (Komen et al. 1994; Polnikov 2005). Here, it is important only to show that the different frequency representations for functions $In(\omega)$ and $Dis(\omega)$ may lead to different forms of a stationary part in the wind-wave spectrum. This expected result is of fundamental importance for understanding the physics of formation a stationary part of the wind-wave spectrum.

There are several steps in executing the task.

Step 0, initialization. The initial shape of spectrum $S(\omega,\theta)$ is given in the form of typical JONSWAP representation (Komen et al, 1994)

$$S_J(\omega,\theta) = S_{PM}(\omega)\gamma^{[(\omega/\omega_p-1)^2/2\sigma^2]}\Psi(\theta-\theta_u) \quad , \tag{21a}$$

where

$$S_{PM}(\omega) = 0.01 * g^2 \omega^{-5} \exp[-\frac{5}{4}\left(\frac{\omega_p}{\omega}\right)^4] \quad (21b)$$

is the Pierson-Moskowitz's (PM) frequency spectrum, generalized to an arbitrary value of $\omega_p$. All other parameters and notations in (21a, b) are standard ones (Komen et al. 1994).

In this study we use the JONSWAP spectrum, $S_J$, and the angular distribution, $\Psi(\theta)$, taken in form of $\cos^2(\theta - \theta_u)$. Initial peak frequency is $\omega_p(0) = 2$ rad/s. As far as the initial spectral shape has practically no effect on the spectral shape for a long-term solution of the kinetic equation (Komen et al. 1994; Polnikov 1990), it is not necessary to vary the initial conditions.

The simple form of generalized evolution equation, (8), is solved at one point, as the spatial evolution of spectrum is not principal in this task. The solution of equation (8) is performed in dimensional units, following to the author's algorithm (Polnikov 1990), based on the explicit numerical scheme of the first-order of accuracy. Kinetic integral $Nl(\omega,\theta)$ is calculated exactly with the algorithm of paper (Polnikov 1989).

All numerical calculations are executed in the $(\omega,\theta)$-domain defined by the boundaries:

$$[0.64 \leq \omega \leq 7.04 \text{ rad/s}; \; -180^\circ \leq \theta \leq 180^\circ], \quad (22a)$$

with the exponential frequency-grid defined by rations

$$\omega_i = \omega_1 q^{i-1} \quad \text{for} \quad \omega_1 = 0.64 \text{ rad/s}, \; q = 1.05, \text{ and } \; 1 \leq i \leq I = 50, \quad (22b)$$

and the equidistant set of 36 angles with resolution of $\Delta\theta = 10$ degrees. As it was checked during testing calculations for $N$L-term (see above subsection 4.1.1), the used resolution in the $(\omega,\theta)$-domain is fairly good, to get reliable numerical results.

Step 1. First, equation (8) is solved numerically for the certain versions of $In$ and $Dis$ terms on the evolution-time scales of the order of $10^4$-$10^5$ values of the initial peak-period, $\tau_p = 2\pi/\omega_p$.

Step 2. Then, in the tail-range of frequencies, $\omega > 2\omega_p$, the spectral decay-parameter $n$ is determined by means of the least squares method, supposing the power-like spectrum shape (13).

Step 3. After that, functions $In$ and $Dis$ are changed as shown in table 1, and steps 1 and 2 of this algorithm are again carried out.

Step 4. Finally, the impact of mathematical representations of functions $In$ and $Dis$ on parameter $n$ is determined and summarized in table.1.

Step 5. In parallel, the time-history of one-dimensional nonlinear transfer, $Nl(\omega; t)$, and the time-history of the balance between pumping and dissipation, $B(\omega;t) = [In(\omega;t) - Dis(\omega;t)]$, are established. This allows to check the ratio between $Nl(\omega; t)$ and $B(\omega;t)$ at the tail of spectrum, and to check a presence of the inertial interval.

5. **Results and analysis**

5.1. Results

The cases considered are shown in table 1. Runs 1-3 are related to WAM's *In* and *Dis* terns and their modifications, run 4 to MASNUM's terms, and runs 5-11 to Polnikov's terms. Here we remind, that the factor of dimensionless frequency, $u_*\omega/g$, is used to increase the frequency dependence of a proper term. The meanings of the other changes will be pointed below.

Table 1.
Parameters of simulations and final values of the stationary spectrum-tail parameter *n*

| # run | Form of *Dis*-term | Form of *In*-term | n (±3%) |
|---|---|---|---|
| 1 | $Dis_W$ | $In_W$ | 4.7 |
| 2 | $10 u_*\omega/g \, Dis_W$ | $In_W$ | 6.6 |
| 3 | $Dis_W$ | $10 u_*\omega/g \, In_W$ | 3.5 |
| 4 | $Dis_M$ | $In_M$ | 5.2 |
| 5 | $Dis_b$ | $In_b$ | 5.2 |
| 6 | $Dis_b$ | $In_b/6$ | 5.4 |
| 7 | $T(\theta,\theta_u)=1$<br>$T_0 + T(\omega) = 2 + 30\dfrac{\omega}{\omega_p}$ | $In_b$ | 5.8 |
| 8 | $Dis_b/6$ | $In_b$ | 4.1 |
| 9 | $Dis_b$ | $10 u_*\omega/g * In_b$ | 4.3 |
| 10 | $Dis_b/6$ | $10 u_*\omega/g * In_b$ | 3.7 |
| 11 | $T(\omega,\theta,\theta_u)=1$<br>$M(\omega,\theta,\mathbf{U})=0.003$ | $In_b$ | 3.1 |

Quantitative results of simulations are presented in the last column of table 1. From these results it is seen the following.

For the WAM's terms, first of all (run 1), it should be noted a good correspondence of the falling law to the observations mentioned in the introduction, which means a good tuning the parameterizations used in WAM. Herewith, it is seen that a simple increasing the frequency dependence of *Dis*-term (run 2) leads to an increasing the falling law parameter *n*. In opposite, an increasing the frequency dependence of *In*-term (run 3), leads to a decreasing the falling law parameter.

Results for the MASNUM's terms (run 4) are similar to the results for modified WAM's terms (run 2), because the *Dis*-term in MASNUM is a little more intensive with respect to WAM, whilst the *In*-term is more or less similar to the one used in WAM.

The results for the Polnikov's terms (runs 5-11), confirms the previous ones: increasing the influence of Dis-term, leads to increasing *n* (runs 6, 7); increasing the influence of *In*-term (runs 8-11) leads to a decreasing the falling-law parameter. Though, this case has much more details, therefore, it will be considered below in more details separately, for the reason of a more complicated mechanics of the tail formation in such representation of *Dis*-term.

5.2 Analysis

Results for runs 1-4 are shown in figures (2a, b, c; 3a, b; and 4), and for runs 5-11 does in figures (5a ,b, c, 6a, b, c, and 7a, b, c). In any case, we can see establishing the stationary (equilibrium) shape of spectrum, $S_{eq}(\omega)$. From these figures we can draw several inferences.

First, as seen, in the real wind-wave situation (runs 1, 4, 5), there is no any inertial interval in the frequency band (figures 2c, 5c). Thus, there is no possibility to explain any part of the equilibrium range of real wind-wave spectra of form (2) by the Kolmogorov's model.

Nevertheless, in figures 2a, 3a and 4 (line 3), one can distinguish a small range at the beginning of the falling spectrum ($\omega_p < \omega < 2\omega_p$), where the spectrum slope is smaller than one in the other part of tail ($\omega > 2\omega_p$). In our mind, this is a simple consequence of small values for the source function balance in this frequency range. In the range $\omega_p < \omega < 2\omega_p$, this balance has the kind

$$Nl(S_{eq}(\omega)) + B(S_{eq}(\omega)) \approx 0 \qquad \text{(for } \omega < 2\omega_p\text{)} \ . \qquad (23)$$

This corresponds to the necessary condition of existing the stationary tail of the spectrum. In table 1, we show estimations for value *n* , made in the last frequency range, $\omega > 2\omega_p$.

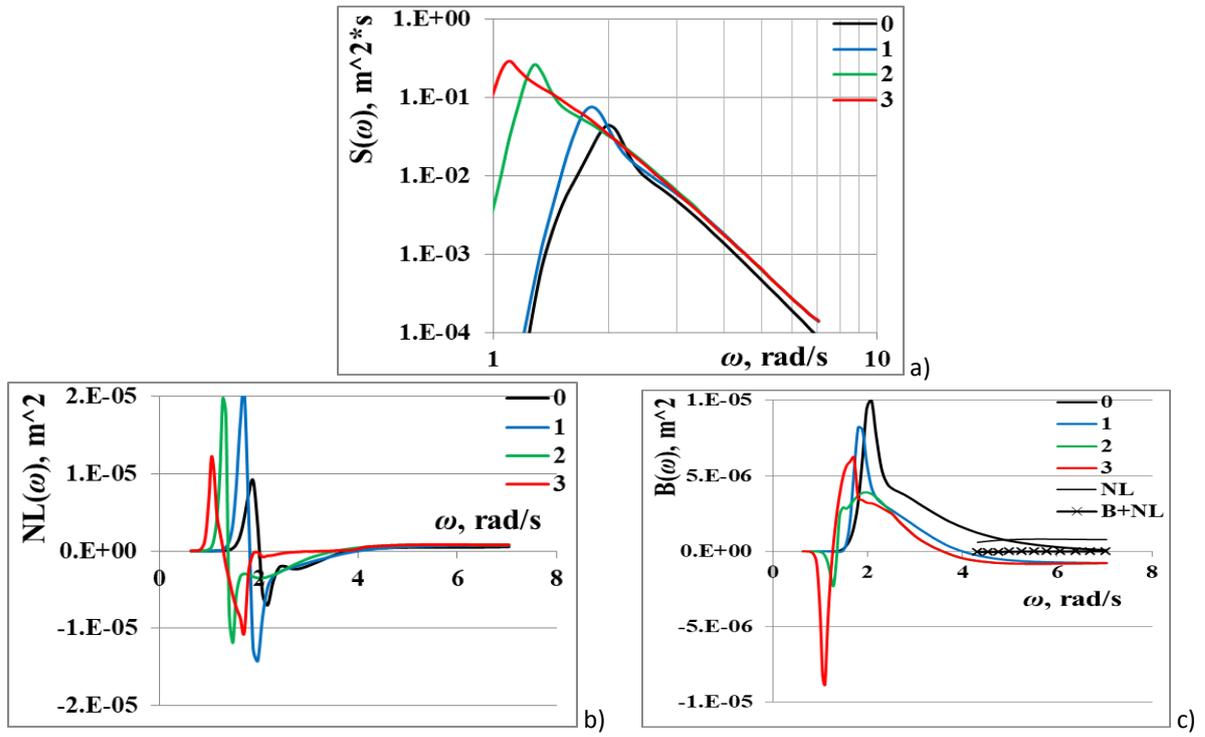

FIGURE 2. Run 1(WAM).  a) Evolution of spectrum $S(\omega,t)$; b) $Nl(\omega,t)$; c) $B(\omega,t)$.
Line 0 corresponds to time  t=0, line 1 to t=3.5*10^3,   line 2 to t=3.8*10^4,line 3 to t=1.1*10^5.
In c), at the end of band, $Nl(\omega)$ and $B(\omega) + Nl(\omega)$ are shown for the stationary range.

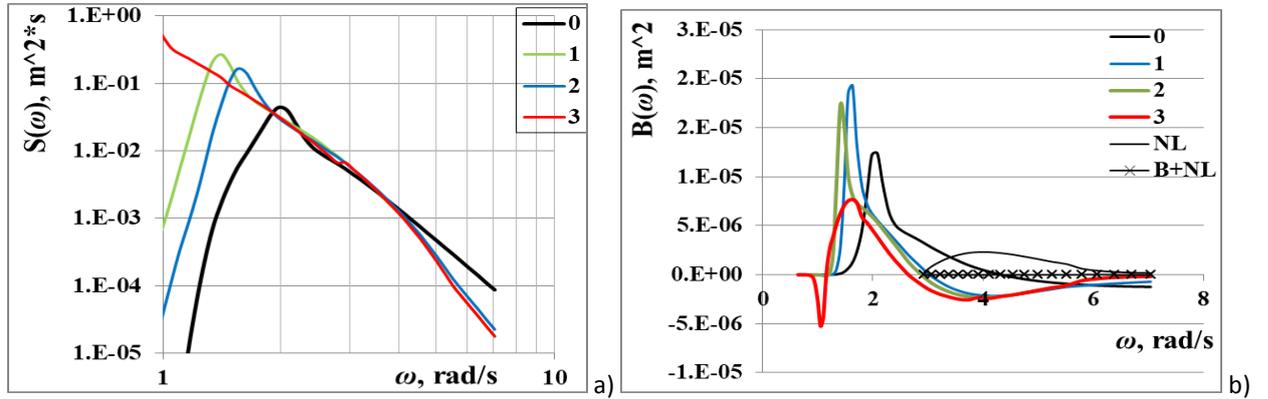

FIGURE 3. Run 2(WAM). a) Evolution of spectrum $S(\omega,t)$ ; b) $B(\omega,t)$.
Line 0 corresponds to time  t=0, line 1 to t=7.7*10^3,  line 2 to t=1.8*10^4, line 3 to t=1.4*10^5.
In b), at the end of band, $Nl(\omega)$ and $B(\omega) + Nl(\omega)$ are shown for the stationary range.

The real zero balance ,

$$Nl(S_{eq}(\omega)) + B(S_{eq}(\omega)) = 0 \qquad , \qquad (24)$$

is shown for runs 1 and 2 separately in figures 2c and 3b,  at the end of frequency band. It is done by adding the lines for $Nl(\omega)$ and the sum $[B(\omega) + Nl(\omega)]$ (for MASNUM it is not shown, as it is very similar to 3b). From these results we can conclude that, according to (24), in the case of *In* and *Dis* terms, linear is spectrum *S*, accepted in the WAM and MASNUM models, one cannot pre-calculate the shape of $S_{eq}(\omega)$ in advance. Moreover, according to (24),  in the case of linear in spectrum *S* dependences for *In* and *Dis* terms,  the final shape of $S_{eq}(\omega)$ is not obligatory to be of power-like form (13), in a wide range of frequency band beyond of $\omega_p$. This inference is fully confirmed in figures 2a, 3a and 4.

Note that in such a case, when one cannot pre-calculate the expected shape of $S_{eq}(\omega)$, there is no way to ensure that $S_{eq}(\omega)$ should be of power-like type. In practice, they could distinguish two ranges of spectral slope. This numerical result could explain the experimental and numerical results of (Tamura et al. 2014).

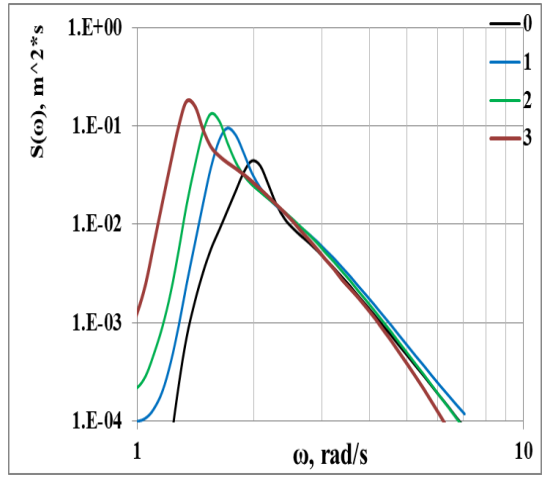

FIGURE 4. Run 4(MASNUM). Evolution of spectrum $S(\omega,t)$ ; b) $B(\omega,t)$.
Line 0 corresponds to time t=0, line 1 to t=4.3*10^3, line 2 to t=9.3*10^3, line 3 to t=2.6*10^4.

5.3. Results and analysis in the case of quadratic *Dis*-term.

In the basic run 5, the stationary range of spectra has a decay law corresponding to $n = 5.2$. This value is very close to $n = 5$, what was theoretically proposed in (Polnikov 2012) during construction of basic function $Dis_b$. Moreover, the stationary shape of spectra is very close to the typical JONSWAP spectrum (figure 5a), what confirms a quality of the basic parameterizations for $In_b$ and $Dis_b$ terms. As above, no inertial interval is seen in fig 5c.

In run 6, the decreasing influence of *In*-term leads to the increase of power $n$ for the stationary tail of wave spectrum (figure 6a), despite of presence the same exact *Nl*-term in the r.h.s. of equation (8). This situation is enforced in run 7, when the frequency dependence of *Dis*-term is exaggerated (not shown). Just in this case, the falling law is very close to form "$\omega^{-6}$", what is in a full accordance with the zero value for balance function $B(S)$.

In all these runs (5, 6, 7), as it is seen in figures 5c, 6c), the zero balance given by the ratio

$$[In(S_{eq}) - Dis(S_{eq})]|_{tail} = 0, \qquad (25)$$

takes place at the tail range, $\omega > 2\omega_p$. Thus, in the case of quadratic *Dis*-term, it is balance (25) between input and dissipation terms is responsible for the formation of spectral shape in the stationary range of wind-wave spectra, $S_{eq}(\omega)$, when they have the falling parameter $n \geq 5$.

Herewith, according to the general stationary spectrum condition, (24), in this cases, the value of $Nl(S_{eq}(\omega))$ is also very small (as seen in figures 5b, 6b) in the whole stationary range, $Nl(S_{eq}(\omega)) \approx 0$. But it does not mean the existence of Zakharov's solution (4) (Zakharov & Filonenko, 1966), as $Nl$-term does not play any role in formation of $S_{eq}(\omega)$, and $S_{eq}(\omega)$ is defined by the ratio (25), only.

On opposite, in run 8, when the influence of $Dis$-term is decreased, the stationary range has the falling law $n = 4.1$. It is very similar to the pure nonlinear evolution mentioned in subsection 4.1. But in this case, the physics of spectrum-equilibrium is absolutely different, as far as the value $Nl(S_{eq}(\omega))$ is not equal to 0 (see figures 7b, c, and explanation below).

The similar, slow falling laws of the stationary spectrum range are established for runs 9-11. In all these cases (runs 8-11), when the falling-law parameters for $S_{eq}(\omega)$ are n < 5, it occurs that the $Nl$-term plays a very important role, providing the exact balance for the total source function, given by ratio (24). But herewith, $Nl$-term is not zero (see figures 7b, c), and its role differs radically from the statements made in Zakharov & Badulin (2011). In these cases, $Nl$-term does not suppress other terms, but supports them, ensuring for making balance (24).

=

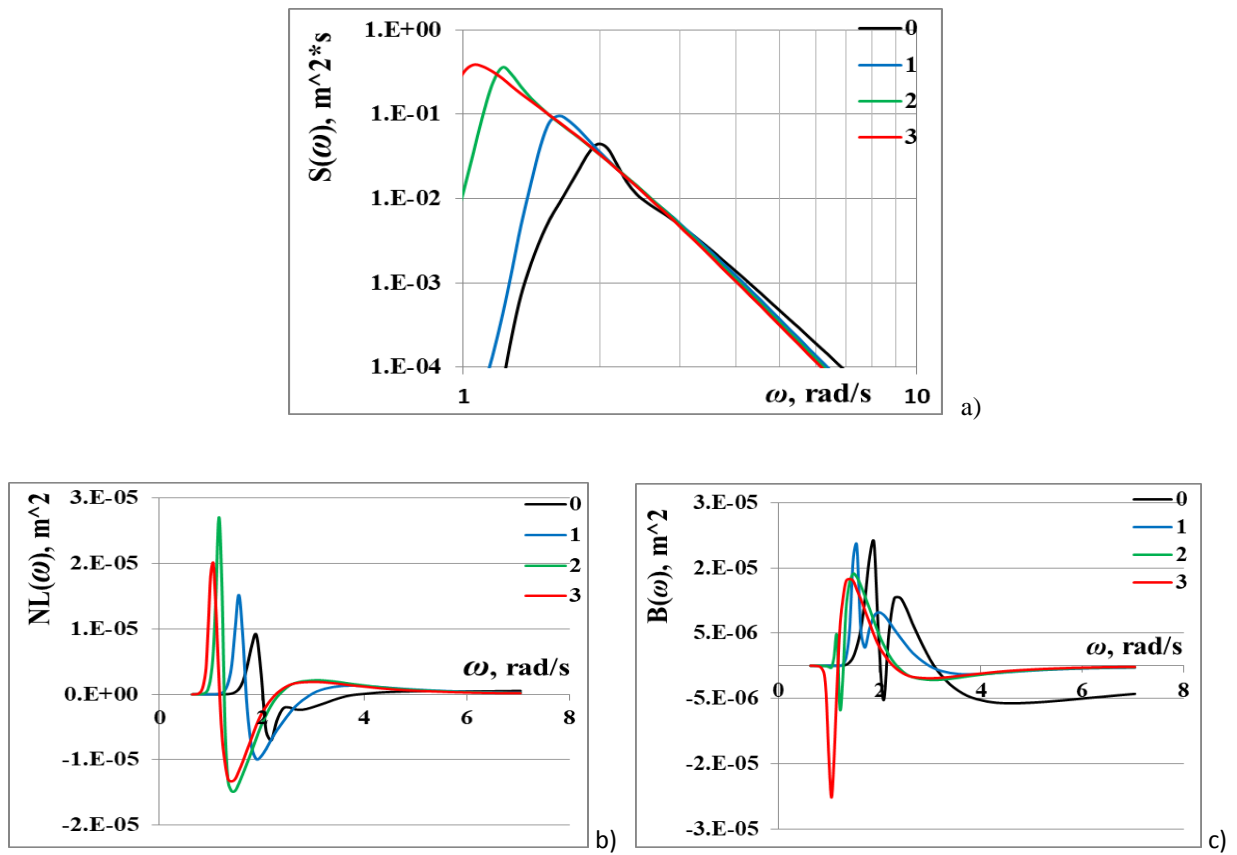

FIGURE 5. Run 5(Polnikov' terms). a) Evolution of spectrum $S(\omega,t)$; b) $Nl(\omega,t)$; c) $B(\omega,t)$.
Line 0 corresponds to time t=0, line 1 to t=3.1*10^3, line 2 to t=2.4*10^4, line 3 to t=0.8*10^5.

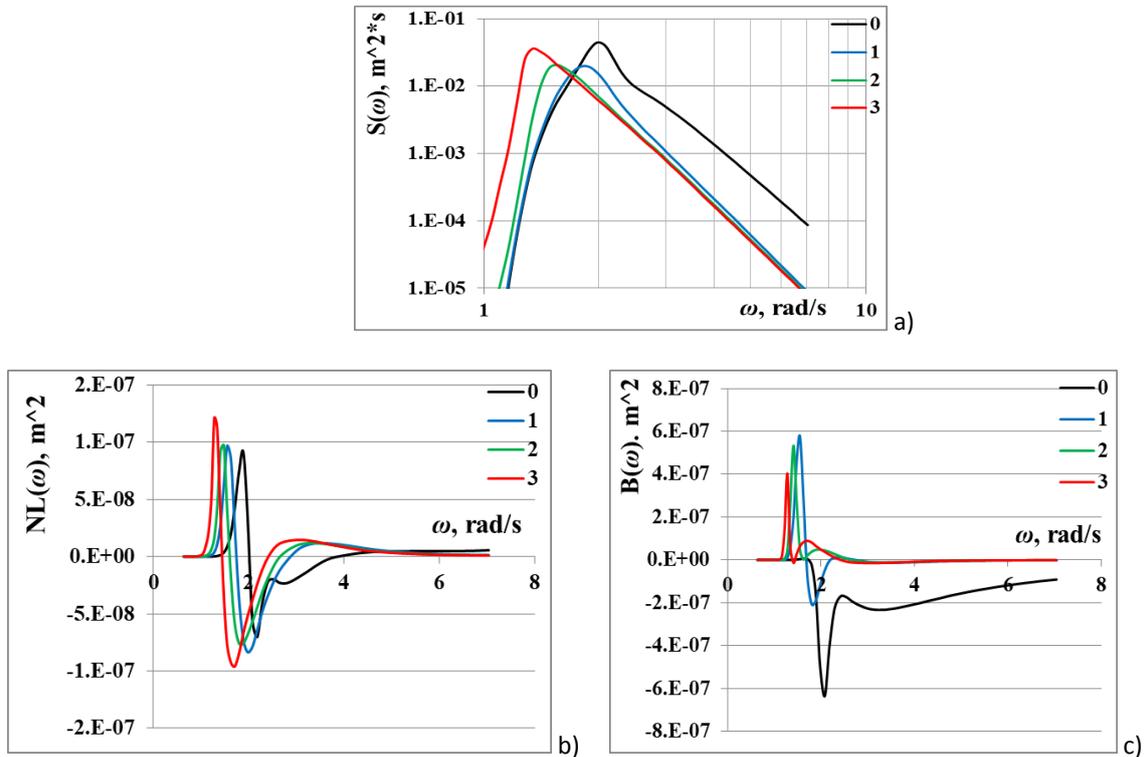

FIGURE 6. Run 6. a) Evolution of spectrum $S(\omega,t)$ ; b) $Nl(\omega,t)$; c) $B(\omega,t)$.
Line 0 corresponds to time t=0, line 1 to t=1.1*10^4, line 2 to t=3.6*10^4, line 3 to t=1.1*10^5.
In b) and c), line 0 is reduced by factor 100 for comparability.

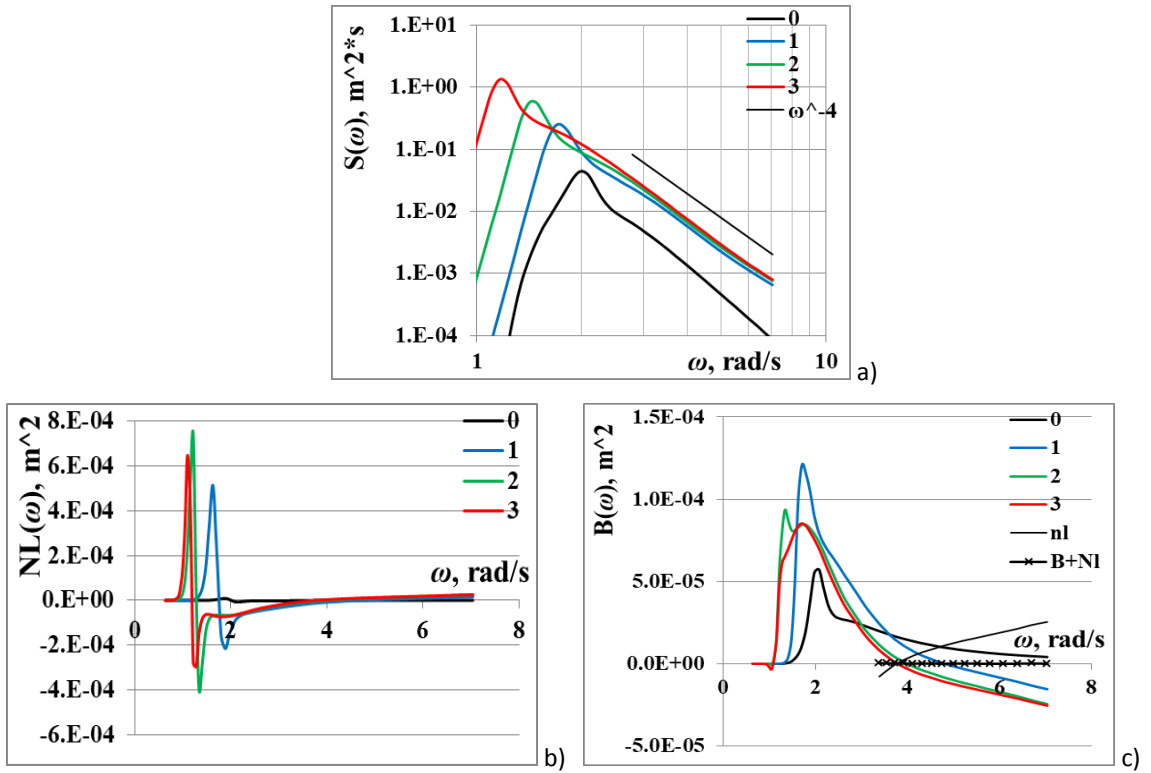

FIGURE 7. Run 8. a) Evolution of spectrum $S(\omega,t)$ ; b) $Nl(\omega,t)$; c) $B(\omega,t)$.
Line 0 corresponds to time t=0, line 1 to t=1.3*10^3, line 2 to t=3.7*10^3, line 3 to t=5.1*10^4.
In c), at the end of band, $Nl(\omega)$ and $B(\omega)+ Nl(\omega)$ are shown for the stationary range.

Main inferences for the case of nonlinear in *S* representation of *Dis*-term are as follows.
Results found show that the variations of intensity and frequency dependences for functions

*In(ω,S)* and *Dis(ω,S)* have a strong direct impact on formation the stationary spectrum-tail, despite of existing exact nonlinear-term *Nl(ω,θ)* in the source function. Consequently, the assumption of Zakharov & Badulin (2011) about adequacy of shearing the term *Nl(ω)* on a sum of large positive and negative parts (9), which supposedly suppress the influence of terms *In(ω)* and *Dis(ω)*, is not confirmed by the exact calculations. On contrary, it is the mechanisms of pumping and dissipation of waves determine a formation of the wind-wave spectrum tail.

Herewith, the role of the nonlinear term is different for different shapes of $S_{eq}(\omega)$. If the value of falling law parameter, *n*, for stationary tail $S_{eq}(\omega)$ is below 5, then the shape of $S_{eq}(\omega)$ is defined by the whole balance of source function, (24). In this case, *Nl*-term is not zero, and plays the role of stabilizing mechanism providing balance (24). This fact is due to positive values of *Nl*-term (at the tail range) for the slow falling spectra (figures 5c, 6c). But if $n \geq 5$ for stationary spectrum $S_{eq}(\omega)$, the shape of latter is defined by balance (25), and *Nl*-term does not play any role on formation of $S_{eq}(\omega)$. In this case, values of *Nl*-term are too small (figures 5b, 6b) to play any role in the equilibrium range formation for wind-wave spectra.

These results are absolutely new numerical findings, which have an important physical meaning in understanding the mechanics of the stationary range formation in the wind wave spectra.

To finish the case of fast falling spectra and the choice of *Dis(S)* nonlinear in *S*, we note that in this case one could estimate in advance the shape of equilibrium spectrum $S_{eq}(\omega)$ from ratio (25). In such a case, one may write analytical representations of terms *In* and *Dis* in the forms

$$In(u_*,\omega,\theta,S) = \alpha(u_*,\omega,\theta)S(\omega,\theta), \quad \text{and} \quad Dis(u_*,\omega,\theta,S) = \gamma(u_*,\omega,\theta)S^2(\omega,\theta). \tag{26}$$

Then, from balance condition (25), one immediately obtains (in the tail range, $\omega > 2\omega_p$) an expected shape of the equilibrium spectrum of the form

$$S_{eq}(u_*,\omega) \approx \alpha(u_*,\omega)/\gamma(u_*,\omega) \qquad (\text{for } \omega > 2\omega_p) \ . \tag{27}$$

The numerical results presented above for runs 5-7 confirm fully the said.

**6. Conclusions and final remarks**

The main conclusions are as follows.

6.1. On the basis of exact solutions of Eq (8) for several known representations of the input and dissipation mechanisms for wind waves, $In(u_*,\omega,\theta,S)$ and $Dis(u_*,\omega,\theta,S)$ (WAM: The WAMDI group 1988; MASNUM: Yang & Yin, 2010; and Polnikov 2005; Samiksha et al. 2015), one can state that the inertial interval does never realize in real wind waves. Therefore, in real wind waves, the Kolmogorov's spectrum model cannot be applied for interpretation of the spectrum-tail formation.

6.2. The ratio between functions $In(u_*,\omega,\theta,S)$ and $Dis(u_*,\omega,\theta,S)$ determines the stationary (equilibrium) range in the wind-wave spectrum, $S_{eq}(u_*,\omega)$, which follows from the zero-balance for the total source function, (24).

Though, mechanics of the tail formation is different for different mathematical representation for functions $In(S)$ and $Dis(S)$, and different falling laws of the equilibrium spectrum. For the falling laws of $S_{eq}(\omega)$ with n < 5, and for representations of $In(S)$ and $Dis(S)$ as linear functions in spectrum $S$, the shape of equilibrium spectrum $S_{eq}(u_*\omega)$ is always governed by the balance for full source function (24). In such a case, the nonlinear mechanism plays a crucial role to ensuring the ratio (24); though, one cannot pre-calculate the shape of $S_{eq}(u_*,\omega)$. When n ≥ 5 for the shape of $S_{eq}(\omega)$, and nonlinear versions of representations $Dis(u_*,\omega,\theta,S)$, equilibrium range $S_{eq}(u_*,\omega)$ is determined by more simple ratio (25). In such a case, one can pre-calculate the expected shape of $S_{eq}(u_*,\omega)$ by ratios (26, 27); herewith, the nonlinear mechanism plays no role in formation the shape of $S_{eq}(u_*,\omega)$.

The said means that, in any case, the shape of $S_{eq}(u_*,\omega)$ is always determined by representations of functions $In(u_*,\omega,\theta,S)$ and $Dis(u_*,\omega,\theta,S)$, whilst the role of *NL*-term is subsidiary.

6.3. The variety of decay laws for the equilibrium tail of spectrum, $S_{eq}(u_*,\omega)$, can be explained by a variety of real mechanisms for dissipation of wind waves. As far as the wave-formation conditions realized in the nature are very different, real dependences $In(u_*,\omega,\theta,S)$ and $Dis(u_*,\omega,\theta,S)$ may also differ from their numerical representations usually used in numerical wind-wave models. Therefore, according to formulas (24-27), in the nature, the decay parameter, *n*, may vary significantly for wind-wave spectra observed under different natural conditions. This ensures validity of the said interpretation for all the above empirical results, including both the variance of *n* established in (Liu 1989; Rodrigues & Soares 1999; Young 1998) and the results of Tamura et al. (2014), do not attracting the Kolmogorov's model.

6.4. In addition to the said, formulas (24-27) make it possible to answer the question: why the spectra of form (2) are often observed in small reservoirs (laboratory tanks, lakes, etc.), in contrast to oceanic areas. Apparently, this is due to the small fetches (and times) of the spectra evolution in these cases. Under such conditions, the dissipative processes can have a special dynamics, which is determined by specific mechanisms of the hydrodynamic instability, governing the processes of dissipation. In small reservoirs, this can lead to establishing dissipative processes that differ from those for global water areas.

For example, a small intensity of wave-breaking can lead to a weakening the frequency dependence of dissipation, $Dis(\omega)$, and its intensity (similar to our run 8). As shown in run 8, this leads the weaker law of spectrum decay, often observed in laboratory tanks or lakes. It is also possible that at low evolution times, the decay law of wind-wave spectrum with $n = 4$ is realized only under the influence of pressure pulsations with the "white noise" spectrum. Such a model was analytically considered (neglecting dissipation and nonlinearity) in (Polnikov & Uma 2014).

As seen, all the observed equilibrium spectral shapes $S_{eq}(u_*,\omega)$ can be explained basing on the concept of equation (24) and presentation for the wind-pumping and wave-dissipation mechanisms, without attracting the Kolmogorov's model. 6.5. In the absence of the inertial interval, the role of nonlinear mechanism is reduced mainly to a downshifting the peak frequency. But for slow falling equilibrium spectra, *Nl*-term does play a crucial role of mechanism compensating the input-dissipation misbalance ( $B(\omega;t) \neq 0$) at the equilibrium range of wave spectrum. For the linear representations of *In(S)* and *Dis(S)*, it is the only possibility the get a stable spectrum shape.

Regarding Kolmogorov-type spectra, we can conclude that they are quite realistic, but only under the abstract (academician) conditions, when a fairly wide inertial interval is artificially established in the frequency band, as it was shown in a lot of numerical experiments (Polnikov 1994, 2001, Badulin et al. 2005; among others).

**Acknowledgments**

The author is grateful to academicians G.S. Golitzyn and E.A. Kuznetsov for their comments and remarks at the initial stage of our investigations. This research was jointly supported by the NSFC-Shandong Joint Fund for Marine Science Research Centers under Grant U1606405, the international cooperation project of Indo-Pacific ocean environment variation and air-sea interaction under Grant GASI-IPOVAI-05. F Qiao was supported by International cooperation project on the China-Australia Research Centre for Maritime Engineering of Ministry of Science and Technology, China under grant 2016YFE0101400, Qingdao National Laboratory for Marine Science and Technology through the AoShan Talents Program under grant 2015ASTP and Transparency Program of Pacific Ocean-South China Sea-Indian Ocean under grant 2015ASKJ01.

**References**

V.A. Babanin & Y. P. Soloviev 1998. Field Investigation of Transformation of the Wind Wave Frequency Spectrum with Fetch and the Stage of Development. *J. Phys. Oceanogr*. 28 (4), 563-576.

S. I. Badulin, A. N. Pushkarev, D. Resio & V. E. Zakharov 2005. Self-similarity of wind-driven seas. *Nonlinear Processes in Geophysics*, 12, 891–945.

D.D. Chen-Zhang, C.S. Ruf, F. Ardhuin & J. Park 2016. GNSS-R nonlocal sea state dependencies: Model and empirical verification. *J. Geophys. Res. Oceans*, 121, doi:10.1002/2016JC012308.

M.A. Donelan, J. Hamilton & W.H. Hui 1985. Directional spectra of wind generated waves. *Phil. Trans. R. Soc*. London. A315, 509-562.

K. Hasselmann 1962. On the nonlinear energy transfer in a gravity wave spectrum. P 1. General theory. *J. Fluid Mech*. 12 (4), 481-500.

K. Hasselmann 1974. On the spectral dissipation of ocean wave due white capping. *Boundary Layer Meteorol.* 6 (1), 107-127.

K. Hasselmann, T.P. Barnett, E. Bouws, E. et al., 1973. Measurements of wind-wave growth and swell decay during the Joint North Sea Wave Project (JONSWAP). *Dtsch. Hydrogr. Z.* 8 (12). 95p.

K.K. Kahma & C.J. Calkoen 1992. Reconciling discrepancies in the observed growth of wind-generated waves. *J. Phys. Oceanogr*. 22, 1389–1405.

S.A. Kitaigorodskii 1962. Applications of the theory of similarity to the analysis of wind-generated wave motion as a stochastic process. *Izv. Acad. Sci. USSR, Geophys. Ser*. 1, 105–117.

S.A. Kitaigorodskii 1983. On the theory of the equilibrium range in the spectrum of wind-generated gravity waves. *J. Phys. Oceanogr*. 13(5), 816-827.

S.A. Kitaigorodskii, V.P. Krasitskii & M.M. Zaslavskii 1975. On the Phillips' Theory of Equilibrium Range in the Spectra of Wind-generated Gravity Waves. *J. Phys. Oceanogr*. 5(2), 410-420..

K. Komatsu & A. Masuda. 1996. A New Scheme of Nonlinear Energy Transfer among Wind Waves: RIAM Method—Algorithm and Performance, *J. Oceanology*, 52, 509–537.

G.J. Komen, L. Cavaleri, M. Donelan et al. 1994. Dynamics and Modeling of Ocean Waves. Cambridge University Press, Cambridge. 532 p.

A.A. Kubryakov, V.G. Polnikov, F.A. Pogarskii & S.V. Stanichnyi 2016. Comparing Numerical and Satellite Data of Wind Wave Fields in the Indian Ocean. *Russian Meteorology and Hydrology* 41, 130–135.

P.C. Liu 1989. On the slope of the equilibrium range in the frequency spectrum of wind waves. *J. Geophys. Res.* 94 (C4), 5017-5023.

A. Masuda 1980. Nonlinear Energy Transfer between Wind Waves, *J. Phys. Oceanogr.*, 10(12). 2082–2093,

J.W. Miles 1957. On the generation of surface waves by shear flows. Pt. 1 . *J. Fluid Mech*. 3(2), 185-204.

A.S. Monin & A.M. Yaglom 1971. Statistical Fluid Mechanics: Mechanics of Turbulence, v. 2. The MIT Press, Cambridge, 720 p.

O.M. Phillips 1958. The equilibrium range in the spectrum of wind-generated waves. *J. Fluid Mech.* 4, 231-245.

O.M. Phillips 1985. Spectral and statistical properties of the equilibrium range in wind-generated gravity waves. *J. Fluid Mech.* 156, 505-631.

W.J. Pierson & L. Moskowitz 1964. A proposed spectrum for fully developed wind seas based on the similarity theory of S.A. Kitaigorodskii. *J. Geophys. Res*. 69(24), 5181-5190.

V.G. Polnikov 1989. Calculation of the nonlinear energy transfer through the surface gravity wave spectrum. *Izv. Acad. Sci., Atmos. Ocean Phys*. 25(11), 118-123.

V.G. Polnikov 1990. Numerical Solution of the kinetic equation for surface gravity waves, *Izv. Acad. Sci., Atmos. Ocean. Phys.* 26(2), 118–123.

V.G. Polnikov 1994. Numerical modeling of flux spectra formation for surface gravity waves. *J. Fluid Mech.* 278, 289-296.

V.G. Polnikov 2001. Numerical modeling of the constant flux spectra formation for surface gravity waves in a case of angular anisotropy. Wave Motion 33, 271-282.

V.G. Polnikov 2005. Wind-Wave Model with an Optimized Source Function. Izvestiya, Atmos. and Ocean. Phys. 41, 594–610.

V.G. Polnikov 2010. Numerical Modelling of Wind Waves. Problems, Solutions, Verifications, and Applications. In: Horizons in World Physics. Editor: Albert Reimer. Nova Science Publishers, Inc. Volume 271. Chapter 1. p. 1-69.

V.G. Polnikov 2012. Spectral Description of the Dissipation Mechanism for Wind Waves. Eddy Viscosity Model. *Marine Science* 2(3), 13-26. DOI: 10.5923/j.ms.20120203.01.

V.G. Polnikov & V. Innocentini 2008. Comparative Study of Performance of Wind Wave Model: WAVEWATCH—Modified by New Source Function. *Eng. Appl. Comput. Fluid Mech*. 2(4), 466–481.

V.G. Polnikov & G. Uma 2014. On the Kolmogorov Spectra in the Field of Nonlinear Wind Waves. *Marine Science* 4(3), 59-63. DOI: 10.5923/j.ms.20140403.01.

F. Qiao, Y. Yuan, Y. Yang, Q. Zheng, C. Xia, & J. Ma 2004. Wave-induced mixing in the upper ocean: Distribution and application to a global ocean circulation model, *Geophys. Res. Lett*. 31, L11303: 1-4, doi: 10.1029/2004GL019824.

G. Rodrigues & C.G. Soares 1999. Uncertainty in the estimation of the slope of the high frequency tail of wave spectra. *Applied Ocean Research* 21, 207-213.

D.T. Resio & W. Perrie 1991. A numerical study of nonlinear energy fluxes due to wave-wave interactions. Part 1: Methodology and basic results. J. Fluid Mech. 223, 609– 629.

D.T. Resio, C.E. Long & C.L. Vincent 2004. Equilibrium-range constant in wind-generated wave spectra. *J. Geophys. Res*. 109, C01018, doi:10.1029/2003JC001788.

D.T. Resio & W. Perrie 1991. A numerical study of nonlinear energy fluxes due to wave-wave interactions. Part 1. Methodology and basic results, *J. Fluid Mech*. 223, 609–629.

S.V. Samiksha, V.G. Polnikov, P. Vethamony et al. 2015. Verification of model wave heights with long-term moored buoy data: Application to wave field over the Indian Ocean. *Ocean Engineering.* 104, 469-479.

H. Tamura, W.M. Drennan, E. Sahl_& H.C. Graber 2014. Spectral form and source term balance of short gravity wind waves, *J. Geophys. Res. Oceans*. 119, 7406–7419, doi:10.1002/2014JC009869.

Y. Toba 1972. Local balance in the air-sea boundary processes. Pt. 1: On the growth process of wind waves. *J. Oceanogr. Soc. Japan*. 28(3), 109-121.

H.L. Tolman 2014. Technical Note. User manual and system documentation of WAVEWATCH III, version 4.18. NOAA, National weather service, NCEP. Washington. 311 p.

H.L. Tolman & D.V. Chalikov 1996. Source terms in a third-generation wind wave model. *J. Phys. Oceanogr.* 26 (11), 2497–2518.

The WAMDI group 1988. The WAM model − A third generation ocean wave prediction model. *J. Phys. Oceanogr*. 18, 1775-1810.

The WISE Group 2007. Wave modelling: state-of-the art. *Progress in Oceanography* 75, 603-674. Doi:10.1016/j.pocean. 2007.05.005.

G. van Vledder 2006. The WRT method for the computation of non-linear four-wave interaction in discrete spectral wave model. J. Coastal Engineering 53, 223-242.

L. Yan 1987. Improved Wind Input Source Term for Third Generation Ocean Wave Modeling, Report No. 87-8. Royal Dutch Meteorological Institute.

Y. Yang, F. Qiao, W. Zhao, Y. Teng & Y., Yuan 2005. The development and application of the MASNUM wave numerical model in spherical coordinates (in Chinese). *Acta Oceanol. Sin.*, 27(2), 1-7.

Y. Yang & X. Yin 2010. User Guide for MASNUM-WAM 2.2. First Institute of Oceanography, SOA, Qingdao. China. 29 p.I. R Young 1998. Observations of the spectra of hurricane generated waves. *Ocean Engineering*, 25 (4-5), 261-274.

I.R. Young & G. van Vledder 1993. A review of the central role of nonlinear interactions in wind-wave evolution. *Phil. Trans. Roy. Soc*. London 342, 505-524.Y. Yuan, F. Hua, Z. Pan, & L. Sun 1991. LAGFD-WAM numerical wave model I. Basic physical model. *Acta Oceanologica Sinica*, 10(4), 483-488.

V.E. Zakharov & S.I. Badulin 2011. On energy balance of wind-driven seas. *Doklady Earth Sciences* 440(2), 1440-1444.

V.E. Zakharov & N.N. Filonenko 1966. The energy spectrum for stochastic oscillation of a fluid's surface, *Dokl. Akad. Nauk*. 170, 1292– 1295.

V.E. Zakharov & M.M. Zaslavsky 1982. Kinetic equation and Kolmogorov spectra in weakly turbulent theory of wind waves, *Izv. Atmos. Ocean. Phys.* 18, 747–753.